\documentclass[
 reprint,
 amsmath,
 prl,
 amssymb,
 aps,
 longbibliography,
]{revtex4-1}

\usepackage[pdftex]{graphicx}
\usepackage{wasysym}
\usepackage{amsfonts}
\usepackage{ifsym}
\usepackage{pifont}
\usepackage{footmisc}
\usepackage{soul}
\usepackage{esint}
\usepackage{amsmath}

\makeatletter
\newlength{\apb@width}
\newcommand{\autoparbox}[2][c]{\settowidth{\apb@width}{#2}\parbox[#1]{\apb@width}{#2}}

\makeatother

\newcommand{\namedref}[2]{\hyperref[#2]{#1~\ref*{#2}}}

\newcommand{\tr}{\operatorname{tr}}

\newcommand{\Csphere}{{}^\bullet\kern-1.2pt C}
\newcommand{\Ctorus}{{}^\circ\kern-1.2pt C}

\newcommand{\nn}{\nonumber}

\newcommand{\COMMENT}[1]{}

\newcommand{\neqa}{\nonumber\end{eqnarray}}
\newcommand{\la}[1]{\label{#1}}

\newcommand{\<}{{\langle}}
\renewcommand{\>}{{\rangle}}

\newcommand{\re}{\relax{\rm I\kern-.18em R}}

\def\su2{{SU(2)}}

\def\[{\left[}
\def\]{\right]}

\def\({\left(}
\def\){\right)}
\def\[{\left[}
\def\]{\right]}

\def\<{\langle}
\def\>{\rangle}

\def\2F1{\,_2{\rm F}_1}

\usepackage[usenames,dvipsnames]{xcolor}
\usepackage{hyperref}
\hypersetup{
    colorlinks,
    linkcolor={red!60!black},
    citecolor={blue!70!black},
    urlcolor={blue!80!black}
}
\usepackage{mathbbol}

\usepackage{cancel}
\usepackage{tikz}
\usepackage{ctable}
\usepackage{booktabs}
\usepackage[caption=false]{subfig}
\newcolumntype{L}[1]{>{\raggedright\let\newline\\\arraybackslash\hspace{0pt}}m{#1}}
\newcolumntype{C}[1]{>{\centering\let\newline\\\arraybackslash\hspace{0pt}}m{#1}}
\newcolumntype{R}[1]{>{\raggedleft\let\newline\\\arraybackslash\hspace{0pt}}m{#1}}
\usepackage{mathtools}

\DeclarePairedDelimiterX\braket[2]{\langle}{\rangle}{#1 \delimsize\vert #2}

\newcommand{\beq}{\begin{equation}}
\newcommand{\eeq}{\end{equation}}
\newcommand{\beqq}{\begin{equation*}}
\newcommand{\eeqq}{\end{equation*}}
\newcommand\beqa{\begin{eqnarray}}
\newcommand\eeqa{\end{eqnarray}}
\newcommand\beqaa{\begin{eqnarray*}}
\newcommand\eeqaa{\end{eqnarray*}}
\newcommand\bea{\begin{array}}
\newcommand\eea{\end{array}}

\begin{document}


\title{What is the simplest holographic matrix model?}

\author{ 
Harish Murali$^{a,b}$, Pedro Vieira$^{a,c}$ 
}
\affiliation{$^{a}$Perimeter Institute for Theoretical Physics, 31 Caroline St N, Waterloo, Ontario N2L 2Y5, Canada}
\affiliation{$^{b}$Department of Physics and Astronomy, University of Waterloo, Waterloo, Ontario, N2L 3G1, Canada}
\affiliation{$^{c}$Instituto de F\'isica Te\'orica, UNESP, ICTP South American Institute for Fundamental Research, Rua Dr Bento Teobaldo Ferraz 271, 01140-070, S\~ao Paulo, Brazil}


\begin{abstract}

We study a generalization of a two-matrix model proposed by J. Hoppe in '89. At strong coupling and large $N$, we unveil an emergent area preserving diffeomorphism symmetry of this model, which allows us to completely solve it at leading order. The solution for all $n$-point connected correlators takes the form of a Witten diagram-like expansion in an emergent two-dimensional dual spacetime. We conclude with speculations about the quantization of this dual model and comment on its extensions to higher dimensions and relations to holography more broadly.

\end{abstract}

\maketitle

\section{Introduction} 

In holography, gravity is in the matrix. The degrees of freedom of strongly coupled field theories of large matrices should somehow reorganize themselves as gravitational theories of sorts describing geometry of an emergent space(-time). This seems to be ubiquitous and yet rarely can we see it at work in a controllable regime. We know the matrices in $\mathcal{N}=4$ SYM must encode precise supersymmetrized Einstein equations in asymptotic $AdS_5 \times S^5$ space -- we can sometimes even check the consequences of this prediction through localization or integrability -- but we really do not know the mechanism for how or why this comes about. 

We suggest here what we believe might be the simplest zero-dimensional matrix model where some of this emergent description can be explicitly derived. The model is given by  (We also use $X_1, X_2$ for $X,Y$ \footnote{Similarly, for integration variables we can use $(x,y)$ or equivalently $(x_1,x_2)=\vec{x}$ and $d^2x\equiv dx dy=dx_1 dx_2$. Finally, $1,2$  are Euclidean directions so upper and lower indices are used interchangeably.})
\beq
\!\!\! Z=\int dX dY \exp\Big(N g\,\tr  [X,Y]^2  - \tfrac N g\,\tr\,{H}(X,Y) \Big) \label{ZgenH}
\eeq
where $X,Y$ are big $N\times N$ Hermitian matrices and the strong coupling limit is the large $g$ limit. In this limit $X$ and $Y$ effectively commute \cite{Berenstein:2008eg,GMV} and we can effectively replace the single trace potential ${H}(X,Y)$ by its fully symmetrized form
\beq
\! : H(x,y): \, \text{ where } :x^2y:=\tfrac{1}{3}(XXY\!+\!XYX\!+\!YXX)\,,  \! \label{hNO}
\eeq
and so on, where $H(x,y)$ is a function with no ordering information. The potential is fully general as~$H(x,y)=-\sum_{n,m}t_{n,m} x^n y^m$. We should stress that this commutativity would be straightforward to explain for large~$g$ if~$N$ was finite since it would follow from the huge suppression of the commutator term in (\ref{ZgenH}) but in the large~$N$~'t~Hooft limit, this is quite a non-trivial fact as highlighted in~\cite{GMV,Martina:2026qxg}. Also, the precise scaling in $g$ of the two terms in (\ref{ZgenH}) matters, see appendix \ref{appA}. We call this model the generalized Hoppe model; for $H(x,y)=x^2+y^2$ this is the model proposed by Hoppe in 1989 in \cite{Hoppe} and further solved beyond large $N$ in \cite{KKN}, see also \cite{Vescovi:2024fwt}.

Connected correlators in this matrix model can then be obtained by taking derivatives of (\ref{ZgenH}) with respect to these times $t_{n,m}$ (dropping obvious $g$, $N$ factors)
\beqa
\< \tr(X^2)\>_c &=& \frac{\partial}{\partial t_{20}}\,\log Z \,, \label{higherPt}\\
\< \tr(XY)\tr(XY)\tr(X^2)\>_c &=& \left(\frac{\partial}{\partial t_{11}}\right)^2 \frac{\partial}{\partial t_{20}}\,\log Z \,.  \nn
\eeqa
and so on. 
Note that the second line can be obtained by taking two more derivatives of the first line. More generally, we can obtain any higher point correlator by taking derivatives of one point correlators which we denote as disk amplitudes. These in turn can be obtained through 
\beqa
\tfrac1N\<\tr(X^n Y^m)\>_c = \int dx\, dy\, \rho(x,y) \,x^n y^m \label{disk}
\eeqa
where $\rho(x,y)$ is the joint eigenvalue distribution of the matrices $X$ and $Y$, an object which we can define precisely because we are in the strong coupling regime where these matrices effectively commute. 

As explained in section \ref{secDer}, this model admits a powerful hidden symmetry by area preserving diffeomorphisms which allows us to completely determine this one-cut density for any potential $H$ as simply 
\beq
\rho(x,y)=\frac1{\pi^2} \sqrt{\Lambda - H(x,y)}  \label{rhoMain}
\eeq
where the constant $\Lambda$ is fixed by the normalization condition  $\int dx dy \,\rho(x,y) =1$, and the support is understood to be the region $H(x,y)\leq \Lambda$ \footnote{
A two dimensional droplet with an area preserving symmetry is of course
reminiscent of the $w_\infty$ structure of the $c=1$ matrix model
\cite{Avan:1991kq,Das:1991qb,Dhar:1992hr} and of the $\frac12$-BPS droplets
\cite{Lin:2004nb}. There, however, the plane is a single particle phase space
introduced from the outset; here it is the joint eigenvalue plane of two
matrices, and neither it nor its symmetry is visible in (\ref{ZgenH}).}.  (For $H(x,y)=y^2+V(x)$ this follows from \cite{GMV}, see appendix \ref{appA}.) 

This is it; given the general density (\ref{rhoMain}) we can compute any disk correlator as~(\ref{disk}) and then straightforwardly obtain any higher point correlator as in~(\ref{higherPt}) since taking derivatives with respect to the times is now trivial to do as the density is a manifestly explicit function of these times. This is why we denote this generalized Hoppe model as the \textit{simplest matrix model}. We further append the \textit{holographic} descriptive because the solution we just described admits a compact diagrammatic representation reminiscent of a Witten diagram expansion in the two dimensional space of disk topology given by $H(x,y)\le \Lambda$ as explained in section \ref{secApplication}. We originally arrived at the connected correlators by a much less direct route, 
through painfully long brute-force  combinatorics reviewed in appendix \ref{Brute}. The surprising simplicity of the final answers 
eventually led us to discover the strong coupling area preserving symmetry,
to which we turn in the next section. It should not be confused with the more
conventional relation between large $N$ $U(N)$ and area preserving
diffeomorphisms, also introduced by Hoppe; see
appendix~\ref{ClarificationAppendic}.

\section{ \label{secDer} An Infinite Dimensional Symmetry}
We consider a loop equation; this is nothing but integrating a total derivative so that we have  
\beq
0=\int d^2 X \left(\frac{\partial}{\partial X_i}\right)^{\!a}_{\,\,b} \Big[ (V_i)^b_{\,\,a} \exp(-N S(X_i))  \Big] \label{zero}
\eeq
where $V_i(X_i)$ is a vector field corresponding to an infinitesimal  change of variables 
\beq
\delta X_i=V_i \,.
\eeq
This matrix valued vector field depends on the two matrices, $V_i=V_i(X_1,X_2)$, and it is therefore important to specify how we order the matrices in this definition. We follow the very same kind of fully symmetrized definition introduced above in (\ref{hNO}) so that 
\beq
V_i(X_1,X_2) = : v_i(x_1,x_2) :
\eeq
Note that $V_i$ is a matrix but $v_i$ is a simple function. 

We consider a very special class of transformations corresponding to divergence-free vector fields
\beq
\partial_i v^i = 0 \,.  \label{div} 
\eeq
Such vector fields generate area-preserving diffeomorphisms. The derivative in (\ref{zero}) will generate three terms since it can act on $V$, on the commutator term in $S$ and on the potential $H$ term in $S$. The first two terms beautifully vanish as a consequence of the area preserving condition~(\ref{div}) in this two matrix setup. 
The details -- which we find very instructive -- are in appendix \ref{WeylAp}. 
In the end, we are thus left with the action of the derivative on the potential term which can be further simplified to (details also in the appendix), 
\beq
\< \tr : v^i : : \partial_i H: \> = 0 \,. \label{product}
\eeq
At this point everything is exact and we did not take any limit as long as $H(X_1,X_2)$ is fully symmetrized in (\ref{ZgenH}). 

At strong coupling, the order of the matrices becomes immaterial to leading order so that (\ref{product}) can be simplified further into our most consequential relation 
\beq
\< \tr : v^i \partial_i H: \> = 0 \,. \label{symmetry}
\eeq
This equation is very special -- it is a \textit{symmetry}. In contrast, the loop equation (\ref{zero}) is the trivial statement that we can change integration variables: for any $V$ the change of the measure is compensated by the change of the action. Generically both changes are complicated and all we learn is a relation among the correlators of a single theory. The vanishing of the first two terms above says that the only effect of a divergence-free flow is to replace $H$ by $H + v^i \partial_i H$. The identity then equates two \textit{different} theories: for any area preserving diffeomorphism $\varphi$, with $(H\circ\varphi)(x,y)=H(\varphi_x(x,y),\varphi_y(x,y))$, the free energy $\mathcal F \equiv -\frac g{N^2}\log Z$ obeys
\beq
\mathcal F_H=\mathcal F_{H\circ\varphi} \label{FSym}\,.
\eeq
This implies a similar relation for the joint density
\beq
\rho_{H\circ\varphi} = \rho_H \circ \varphi \,.
\eeq
This follows from deforming $H\to H+\delta H$ in \eqref{FSym} which implies $\int d^2x\, \left(\rho_H(x) - \rho_{H\circ\varphi}(\varphi^{-1}(x))\right) \delta H(x) = 0$ for all variations $\delta H$. A generic $H(x,y)$ can be obtained from $H_0=Y^2+h(X)$ by an area preserving diffeomorphism as illustrated in figure \ref{figArea} so that if we know the density for $H_0$ (which we do \cite{GMV}, see appendix \ref{appA}) we can get the density for any $H$ and this is how we obtain our main result (\ref{rhoMain}). 

 \begin{figure}[t]
     \centering
     \includegraphics[width=\columnwidth]{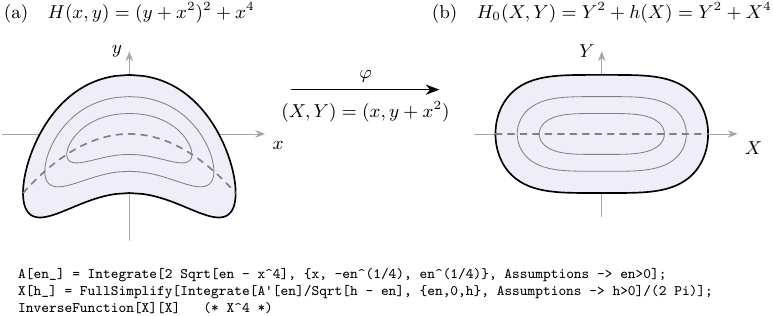}
        \caption{A simple $H(x,y)$ is transformed by an area preserving diffeomorphism into $H_0(X,Y)=Y^2+h(X)$. Here $H(x,y)=(y+x^2)^2+x^4$ and $h(X)=X^4$, and the change of variables $(X,Y)=(x,y+x^2)$ is trivial to find. It clearly maps $H$ into $H_0$ and is clearly area preserving since $dX\wedge dY=dx\wedge dy$. For a more general example we could construct $h(X)$ by imposing that it is such that the given area function $A(E)\equiv\int\Theta\big(E-H(x,y)\big)\,dx\,dy$ would be the same as the transformed expression $\int\Theta\big(E-Y^2-h(X)\big)\,dX\,dY=
        2\int
        dX\,\sqrt{E-h(X)}$. Equating these two functions leads to an explicit solution for $h(X)$ or -- more precisely -- for its inverse $X(h)$, defined for even $h$, in terms of the so-called Abel inversion formula $2\pi X(h)=\int_0^h dE\,A'(E)/\sqrt{h-E}$. The Mathematica code checks this for the example shown. We assume throughout that $H$ has a single minimum and is otherwise sufficiently monotonic; disjoint distributions are treated similarly as in \cite{GMV}.
           %
           }
     \label{figArea}
 \end{figure}

\section{\label{secApplication} Feynman Diagrams and the Bulk} 
The strong coupling free energy of the generalized Hoppe model is given by 
\begin{equation}    
    \mathcal{F}[H]
    =
    \underset{\Lambda}{\operatorname{ext}}
    \left\{
    \Lambda
    -
    \frac{2}{3\pi^2}
    \int d^2x\,
    \Big(\Lambda-H(\vec{x})\Big)^{\tfrac{3}{2}}
    \right\}\,,
\label{eq:HoppeExplicitLeadingFreeEnergy}
\end{equation}
up to an $H$ independent constant. 
The integral is restricted to the region where $H(x,y)\leq \Lambda$. Equivalently, 
\begin{equation}
    \mathcal F[H]
    =
    \underset{\substack{\rho\geq0\\ \int d^2x\,\rho=1}}
    {\operatorname{min}}
    \left\{
    \int d^2x\,\Big(H\rho
    +
    \frac{\pi^4}{3} \rho^3 \Big)
    \right\}.
\label{eq:HoppeLeadingDensityFunctional}
\end{equation}
The correctness (and equivalence) of these expressions follows from the fact that they both lead to the required relation 
\begin{equation}
    \delta\mathcal F[H]
    =
    \int d^2x\,\rho(\vec{x})\,\delta H(\vec{x}),
\end{equation}
with density given by (\ref{rhoMain}), as one can easily check \footnote{The check is particularly trivial since we only need to consider the explicit variation of these expressions with respect to $H$ since the induced variation of the other variables vanishes trivially by the equations of motion. We reproduce (\ref{rhoMain}) from the second relation (\ref{eq:HoppeLeadingDensityFunctional}) when we add the normalization constraint $\int \rho=1$ to the action using a Lagrange multiplier $\Lambda$.}. 

We can think of (\ref{eq:HoppeLeadingDensityFunctional}) or (\ref{eq:HoppeExplicitLeadingFreeEnergy}) as on-shell actions for a dual theory living in the two dimensional space $(x,y)$ of disk topology. From these actions -- or from the one point functions (\ref{disk}) we can then compute any correlators by taking variations with respect to $H$. 

To compute connected correlators, we choose $H=H_0(\vec{x})-J(\vec{x})$ for some background $H_0$ and where the source $J=\sum t_n f_n(\vec{x})$ with times~$t_n$ (set to zero at the end) sources single trace operators~$\tr f_n(X,Y)$ (at strong coupling the order of $X,Y$ in the single trace is immaterial and so to each operator we have the corresponding function $f_n(\vec{x})$). The density for $H$ and for $H_0$ are both normalized and thus 
\beq
0= \int d^2 x\, \rho(\lambda+J,\vec{x})-\int d^2 x \,\rho(0,\vec{x})
\eeq
where we highlighted that the density only depends on the normalization constant $\Lambda\equiv \Lambda_0+\lambda$ and on the source $J$ through the simple combination $\lambda+J$, see (\ref{rhoMain}). This is of critical importance of course and is the main consequence of the area-preserving diffeo symmetry unveiled in the previous section.
Expanding this equation for small sources then yields  
\begin{equation}
    0
    =
    \sum_{m\geq1}\frac1{m!}
    \int d^2x\,\,
    \omega_{m+1}(\vec{x})
    \bigl(\lambda+J(\vec{x})\bigr)^m.
\label{eq:HoppeGeneralNLambdaRecursion}
\end{equation}
where $\omega_m$ are simply $m-1$ derivatives of the density for $H_0$ (distributionally speaking -- first integrate moments against $\rho_{H_0}$; then take derivatives and not the other way around -- see appendix \ref{wmomentaAp}). For the vacuum $H_0 = x^2+y^2$, its non-vanishing moments are simply given by
\begin{equation}
    \int d^2x\,
\omega_m^{(0)}(\vec x)|\vec x|^{2s}=\frac{L^{5+2(s-m)}}{2\sqrt\pi}\frac{\Gamma(s+1)}{\Gamma\left(s-m+\frac72\right)}\,.
\label{eq:momentsW}
\end{equation}
for $m=2,3,\dots$ and $s=0,1,2,\dots$ and $L=\left(\frac{3\pi}2\right)^{\frac13}$ is the size of the vacuum density's support (see Appendix \ref{appA}). To simplify the first term in (\ref{eq:HoppeGeneralNLambdaRecursion}) we shift $\lambda \to \lambda+ [J]$ where 
\beq
[J] \equiv \frac1{\int d^2x\, \omega_2(\vec x)}\int d^2 x \, \omega_2(\vec{x}) J(\vec{x}) \,. 
\eeq
Then (\ref{eq:HoppeGeneralNLambdaRecursion}) becomes simply
\begin{equation}
    \lambda 
    =
 -\sum_{m\geq 2}\frac1{m!}\,
    \frac{\displaystyle\int d^2x\,
    \omega_{m+1}(\vec{x})
    \bigl(\lambda+J_h(\vec{x})\bigr)^m}{\int d^2x\, \omega_2(\vec x)}.
\label{eq:HoppeGeneralNLambdaRecursion2}
\end{equation}
where $J_h$ takes the same form as $J$ but now $f_n$ are replaced by their centered expressions $f_n-[f_n] \equiv h_n$. This equation generates a rooted-tree expansion \footnote{Had we not performed the useful shift of $f\to h$ described above, we would still have a nice diagrammatic solution but we would now have tadpole vertices slightly complicating the analysis.}.  Each use of
the right-hand side introduces one contact vertex and one propagator factor of the normalization mode $\lambda$ equal to~$-\kappa$, defined as
\beq
    \kappa \equiv \left(\int d^2x\,\omega_2(\vec x)\right)^{-1}.
\eeq

To get the correlators, single out the first
insertion as explained in the introduction and write
\begin{equation}
\begin{aligned}
    &\frac{N^{n-2}}{g^{n-1}}\left\langle
    \prod_{i=1}^n\tr f_i(X,Y)
    \right\rangle_c
    \\
    &\qquad=
    \left.
    \frac{\partial}{\partial t_2}\cdots
    \frac{\partial}{\partial t_n}
    \int d^2x\,\rho(\lambda+J_h,\vec{x})h_1(\vec{x})
    \right|_{t=0}.
\end{aligned}
\label{eq:HoppeGeneralNRootedDerivative}
\end{equation}
Expanding the density in $\lambda+J_h$, and then using \eqref{eq:HoppeGeneralNLambdaRecursion2} for every occurrence of $\lambda$, generates every tree rooted at the external label $1$.  Forgetting this choice of root gives every tree with the $n$ external labels exactly once. The factors $1/m!$ in (\ref{eq:HoppeGeneralNLambdaRecursion2}) are canceled by the ways in which the labeled source derivatives can enter the $m$ branches, so there are no additional symmetry
factors.
\begin{figure}[t]
    \centering
  \includegraphics[scale=1]{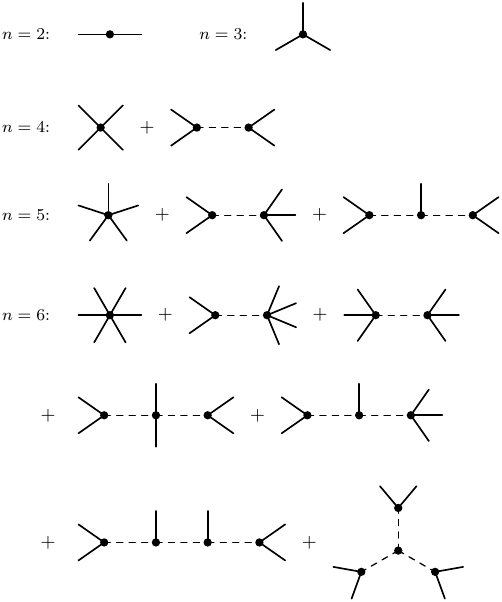}
        \caption{Witten-like diagrams 
       giving all connected correlators of up to six traces following the general expression~(\ref{eq:HoppeGeneralNTreeFormula}). Each solid line is attached to an external trace and the dashed lines come with factors of $-\kappa$. The last graph, for instance, would contribute (up to normalization factors) as
      $-\kappa^3\iiiint d^2 x_4\, \omega_3(\vec{x}_4) \prod_{i=1}^3 d^2 x_i  \,\omega_3(\vec{x}_i) h_{2i-1}(\vec{x}_i)h_{2i}(\vec{x}_i)$
         plus the $14$ other inequivalent permutations of the external legs. This is the first graph where a fully internal vertex shows up. }
     \label{fd_fig}
 \end{figure}
More explicitly, let $\mathcal T_n$ be the set of trees with external leaves
labeled $1,\ldots,n$.  Internal vertices have valence at least three; for
$n=2$ we also include the single two-valent contact vertex.  For an internal
vertex $v$, let $d_v$ be its total valence, including both external leaves
and internal edges, and let $I_v$ be the external labels attached directly
to it.  If $E(T)$ is the number of internal edges, then 
\begin{equation}
\begin{aligned}
    \left\langle
    \prod_{i=1}^n\tr f_i(X,Y)
    \right\rangle_c
    &=\frac{g^{n-1}}{N^{n-2}}
    \sum_{T\in\mathcal T_n}
    (-\kappa)^{E(T)}
    \\
    &\,\,\times
    \prod_{v\in T}
    \int d^2x\,
    \omega_{d_v}(\vec x)
    \prod_{i\in I_v}h_i(\vec x)
    .
\end{aligned}
\label{eq:HoppeGeneralNTreeFormula}
\end{equation}
If no external label is attached directly to a vertex, the product over
$I_v$ is one, see figure \ref{fd_fig}. 

It is useful to summarize what is kinematics and what is dynamics. The structure (\ref{eq:HoppeGeneralNTreeFormula}) follows basically from the area-preserving \textit{symmetry} we unveiled in the previous section. The precise \textit{dynamics} of the undeformed Hoppe model are encoded in the hemisphere density of the vacuum and thus enter only when we use (\ref{eq:momentsW}) to evaluate the moments.

Some correlators are especially simple. For example, in terms of charged scalars $Z,\bar Z \equiv X\pm i Y$,
\beq
\!\!G_n\equiv \left\langle
    \tr \bar Z^{J}\prod_{i=1}^n\tr Z^{J_i}
  \!  \right\rangle_c\,.\label{extremalCorr} 
\eeq
At strong coupling, since $\lambda$ is neutral, only the contact diagram contributes. At weak coupling, the extremal correlator (\ref{extremalCorr}) was conjectured -- \textit{for any $N$ actually} -- in~\cite{Beisert:2002bb,Corley:2001zk,Bergere:2005az}. At large $N$ we have,
\beq
\!\!\!    G_n = \frac{\delta_{J,\sum_i J_i} J!}{N^{n-1}}\times\!\begin{cases} 
\displaystyle\frac{g^n\,L^{2J-2n+3}}{2\sqrt\pi\,\Gamma\big(J-n+\tfrac52\big)}\,,\quad g\gg1,\\
\displaystyle
\frac{g^J}{\Gamma(J-n+2)}\,\prod_{i}J_i\,,\quad g\ll1\,. 
\end{cases}\!\!\!\label{extremalWeakvsStrong}
\eeq 
We can easily cook up other simple examples where only a few diagrams survive at strong coupling. It would be interesting to interpret the differences and similarities in~\eqref{extremalWeakvsStrong} and to study the full weak to strong coupling interpolation (ideally for some more generic non-extremal correlators as well). Matrix Monte Carlo methods~\cite{Martina:2026qxg,Jha,GMV} could be instrumental in checking future results.

\section{\label{secDiscussion} Future Work and Speculations} 

We have seen that area preserving diffeomorphisms are a powerful emergent symmetry at strong coupling and that they allowed us to compute the free energy and connected correlators of the generalized Hoppe model (\ref{ZgenH}) at leading order. What we find most encouraging, however, is not the solution itself but the mechanism behind it. The two dimensional space on which our diagrammatics lives is nowhere to be seen in the definition (\ref{ZgenH}) -- it emerges, together with its diffeomorphism redundancy, from the strong coupling dynamics of the matrices.

Why area preserving diffeomorphisms and not arbitrary ones? In gravity we are used to all diffeomorphisms being redundancies. The resolution is that we are simply working in a particular gauge. Indeed, we can introduce an area form $\mu(\vec{x})\,dx_1\wedge dx_2$ and write
\beq
\mathcal F[H] = \underset{\Lambda}{\operatorname{ext}} \left(\Lambda - \frac{2}{3\pi^2}\int  \mu(\vec{x}) \big(\Lambda - H(\vec{x})\big)_+^{\frac32} \right) \label{eq:covariantF}
\eeq
which is now manifestly invariant under \textit{arbitrary} diffeomorphisms, with $\mu$ transforming as a density and $H$ as a scalar. Our expression (\ref{eq:HoppeExplicitLeadingFreeEnergy}) is nothing but (\ref{eq:covariantF}) gauge fixed to the flat area form $\mu=1$, and the residual transformations preserving that gauge are precisely those generated by divergence free vector fields -- our symmetry~(\ref{symmetry}). Note that in~(\ref{eq:covariantF}) the sources $H$ and the density $\rho$ only ever couple to the density (equivalently to $\sqrt{g}$) and never to the full metric $g_{\mu\nu}$, which is why an area form is all we needed. It would be interesting to identify sources in the matrix model -- perhaps involving commutators and thus derivatives -- which do couple to the full bulk metric. 

Classically, introducing $\mu$ this way is a rather vacuous move: we added one degree of freedom and one gauge redundancy which promptly eats it. Quantum mechanically it need not be, since we would now sum over bulk data and must decide on a measure for that sum. The Hoppe model is a wonderful playground in which to ask such questions concretely. Can we promote these on-shell actions to a full fledged bulk path integral? Should we integrate (\ref{eq:HoppeExplicitLeadingFreeEnergy}), or (\ref{eq:HoppeLeadingDensityFunctional}), or yet another form which happens to agree on-shell? The sharp test is that the resulting quantum theory should reproduce the $1/N^2$ corrections to the correlators, which on the matrix model side are perfectly well defined and computable in principle.

Also interesting is the strong coupling expansion. 
The tree expansion \eqref{eq:HoppeGeneralNTreeFormula} suggests a notion of bulk \textit{locality}. At strong coupling, each trace is labeled by a function $f_i(x,y)$, which we can think of as a wavefunction. Every vertex is local -- all the functions attached to it are evaluated at the same point and integrated against a single $\omega_{d_v}$. Different vertices are connected by the dashed propagators of figure \ref{fd_fig}, and information can therefore pass between different points. At leading order, this propagation is exceptionally simple: every internal line is just a constant $-\kappa$, and carries no information about the separation of its endpoints. The leading strong coupling theory therefore has local interaction vertices connected by a simple propagating mode, but not yet the position-dependent propagators of an ordinary bulk field theory. It would be illuminating to see whether subleading orders in $g$ generate such dependence, and whether genuinely non-perturbative effects are where non point-like objects, such as strings, become visible.
\begin{figure}[t]
    \centering
  \includegraphics[scale=0.3]{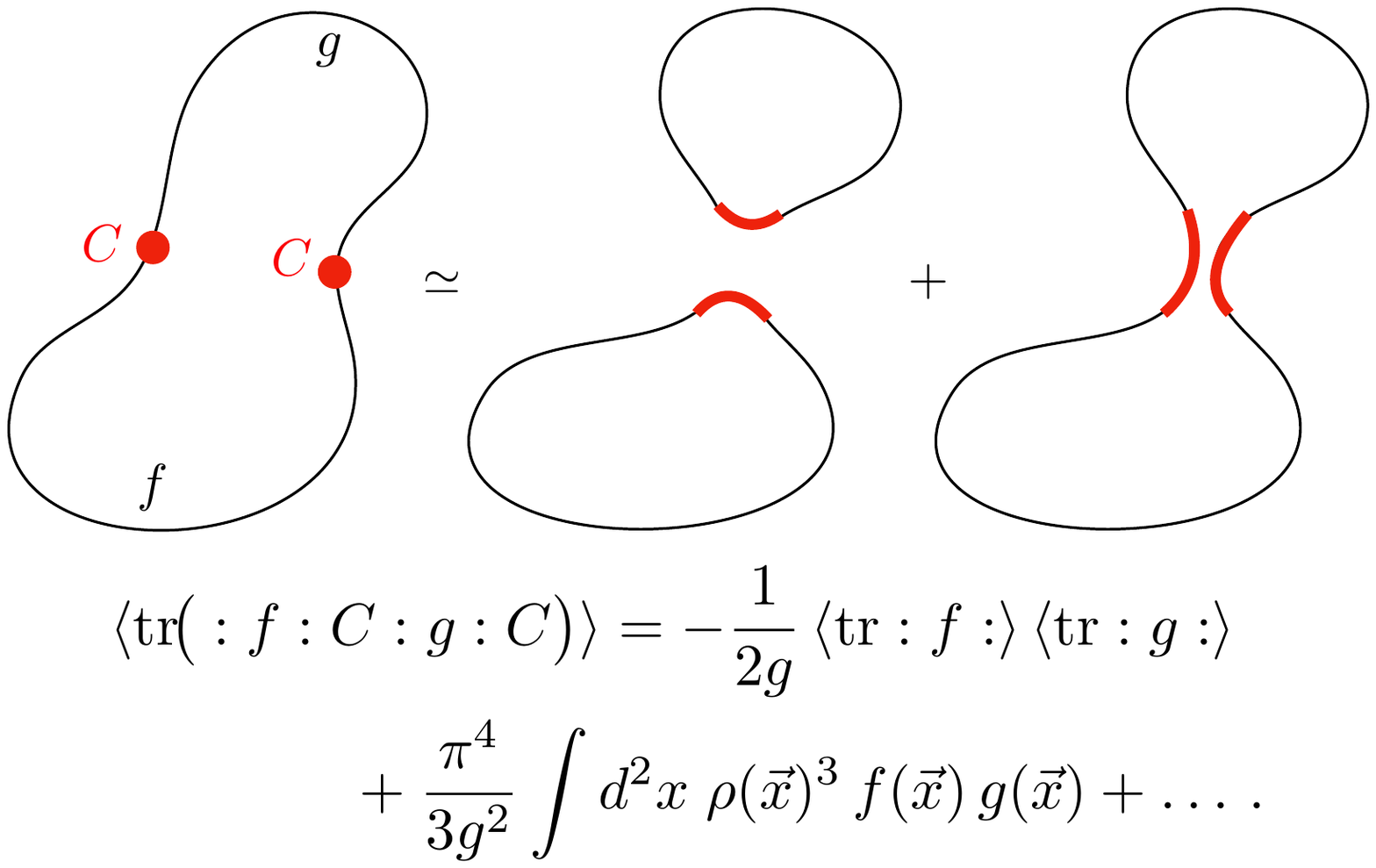}
     \vspace{-1.2cm}
        \caption{The simplest insertion of commutators corresponds to inserting two commutators in a word. 
At leading order, the two commutators simply pair up and cut the trace into two planar faces. The bulk sees two independent points and nothing propagating between them. All the local information therefore hides in the subleading contact term in the figure, which suggests that the good observables are not the single traces themselves but the subtracted combinations\\\\
$
{}\,\,\,\qquad\widehat{\mathcal{O}}[f,g] \equiv \tr\!\big(:f:\, C\, :g:\, C\big) + \frac{1}{2gN}\, \tr :f:\ \tr :g: \,,
$
\\\\
in which the disconnected pairing has been removed by hand. This is reminiscent of how one corrects naive single trace operators into proper single particle operators in $\mathcal N=4$ SYM by carefully subtracting double trace terms \cite{singleP}. 
        }
     \label{2coms}
 \end{figure}

In the same spirit one can insert commutators~$C=[X,Y]$ inside the single traces in order to not have them collapse to a point. Since a commutator is an intrinsically off-diagonal object, one might have expected such insertions to connect two \textit{different} points of the bulk. See figure \ref{2coms} for a first exploration.

It is natural to ask how much of this survives with more matrices. In \cite{Martina:2026qxg} a family of multi-matrix models with $D$ bosonic matrices was found to be commuting at strong coupling, provided the number of fermionic degrees of freedom reaches the critical value $\mathcal N_c=2(D-2)$. The Hoppe model is the $D=2$ member of this family, where no fermions are needed for commutativity. It is therefore natural to ask whether the strong coupling dynamics for these commuting models is also constrained by diffeomorphism invariance, and whether volume preserving ones are special. Since these models contain fermions, one should also look for the Grassmann odd counterpart of our symmetry. The analysis of more holographically established models --  starting with~\cite{Martina:2026qxg} and aiming at~\cite{IKKT,Bonelli,Joao1,Joao2,Sean1} -- would be most welcome. We leave these explorations to future works.

Finally, there exists another remarkable three matrix model generalization which we discuss in more detail in appendix \ref{xyzAp} with the following action
\beq
S_{\text{XYZ}} = -2 \tr [X,Y]Z  + \tfrac 1g \,\tr\,{H}(X,Y,Z) 
\label{ZgenH3} \,.
\eeq
(We learned from Davide Gaiotto that it arises naturally in the B-model formulation of twisted holography~\cite{Costello:2018zrm, Budzik:2023xbr,Gaiotto:2024}; it would be fascinating to understand how this perspective relates to the action functionals and Feynman rules derived here.) This XYZ model has been suggested as a generalization of the Hoppe model already in \cite{Hoppe}, see also discussion in \cite{GMV,OConnor:2012vwc,Filev:2013pza} and indeed, some of the above analysis extends to this model as well. The main result is that the two dimensional area preserving symmetry generalizes to a three dimensional volume preserving symmetry, albeit with important subtleties. (It remains to be seen whether these subtleties are features, bugs or something in between.) Furthermore, for this three matrix model, the density turns out to be uniform inside a three-dimensional droplet \footnote{As explained in Appendix \ref{xyzAp}, this density only captures the moments of symmetrized traces. Earlier literature, including by us \cite{GMV}, incorrectly claims that this is the joint eigenvalue density for general traces.} -- all the dynamics is encoded in the shape of the boundary, analogous to the~$\frac12$-BPS case in $\mathcal N=4$ SYM \cite{Lin:2004nb,Berenstein:2005aa}.

It would be very interesting to understand how the matrix bootstrap \cite{Anderson:2016rcw,Lin:2020mme} interfaces with the area preserving symmetry. Note that (\ref{product}) is already imposed there, being the loop equation (\ref{zero}) for a divergence-free vector field; what is not a loop equation is the strong coupling reordering leading to (\ref{symmetry}). In explorations with Xinran Su and Zechuan Zheng we found that the numerical bootstrap of the Hoppe model in \cite{Kazakov:2021lel} struggles more and more as the coupling grows -- precisely where our analytics simplify. These are probably two faces of the same coin: as the matrices commute, different orderings of a word degenerate and the positivity matrix develops many near-null directions. Clearly there is something important to be understood here.

\begin{acknowledgments}
\section*{Acknowledgments}

We thank Federico Ambrosino, Davide Gaiotto, Andrea Guerrieri,  Henry Lin, Juan Maldacena, Adrien Martina, Joao Penedones, Xinran Su, Xiang Zhao and Zechuan Zheng for many enlightening discussions. We thank Simon Caron-Huot for reminding us of Kontsevich's deformation quantization \cite{KD}; we are digesting this work. Research at the Perimeter Institute is supported in part by the Government of Canada through NSERC and by the Province of Ontario through MRI. This work was additionally supported by the ICTP-SAIFR FAPESP grant 2016/01343-7. 
\end{acknowledgments}


\bibliography{draft} 

@article{Lin:2004nb,
    author = "Lin, Hai and Lunin, Oleg and Maldacena, Juan Martin",
    title = "{Bubbling AdS space and 1/2 BPS geometries}",
    eprint = "hep-th/0409174",
    archivePrefix = "arXiv",
    reportNumber = "PUPT-2136",
    doi = "10.1088/1126-6708/2004/10/025",
    journal = "JHEP",
    volume = "10",
    pages = "025",
    year = "2004"
}

@article{Berenstein:2005aa,
    author = "Berenstein, David",
    title = "{Large N BPS states and emergent quantum gravity}",
    eprint = "hep-th/0507203",
    archivePrefix = "arXiv",
    doi = "10.1088/1126-6708/2006/01/125",
    journal = "JHEP",
    volume = "01",
    pages = "125",
    year = "2006"
}

@article{OConnor:2012vwc,
    author = "O'Connor, Denjoe and Filev, Veselin G.",
    title = "{Near commuting multi-matrix models}",
    eprint = "1212.4818",
    archivePrefix = "arXiv",
    primaryClass = "hep-th",
    reportNumber = "DIAS-STP-12-12",
    doi = "10.1007/JHEP04(2013)144",
    journal = "JHEP",
    volume = "04",
    pages = "144",
    year = "2013"
}

@article{Bergere:2005az,
    author = "Bergere, M. C.",
    editor = "Bertola, M. and Harnad, J.",
    title = "{Correlation functions of complex matrix models}",
    eprint = "hep-th/0511019",
    archivePrefix = "arXiv",
    reportNumber = "SPHT-T05-174",
    doi = "10.1088/0305-4470/39/28/S01",
    journal = "J. Phys. A",
    volume = "39",
    number = "28",
    pages = "8749--8774",
    year = "2006"
}

@article{Corley:2001zk,
    author = "Corley, Steve and Jevicki, Antal and Ramgoolam, Sanjaye",
    title = "{Exact correlators of giant gravitons from dual N=4 SYM theory}",
    eprint = "hep-th/0111222",
    archivePrefix = "arXiv",
    reportNumber = "BROWN-HET-1292",
    doi = "10.4310/ATMP.2001.v5.n4.a6",
    journal = "Adv. Theor. Math. Phys.",
    volume = "5",
    pages = "809--839",
    year = "2002"
}

@article{Martina:2026qxg,
    author = "Martina, Adrien and Murali, Harish",
    title = "{On the strong coupling limit of Yang-Mills matrix models}",
    eprint = "2607.21593",
    archivePrefix = "arXiv",
    primaryClass = "hep-th",
    month = "7",
    year = "2026"
}

@article{Costello:2018zrm,
    author = "Costello, Kevin and Gaiotto, Davide",
    title = "{Twisted holography}",
    eprint = "1812.09257",
    archivePrefix = "arXiv",
    primaryClass = "hep-th",
    doi = "10.1007/JHEP01(2025)087",
    journal = "JHEP",
    volume = "01",
    pages = "087",
    year = "2025"
}

@article{Beisert:2002bb,
    author = "Beisert, N. and Kristjansen, C. and Plefka, J. and Semenoff, G. W. and Staudacher, M.",
    title = "{BMN correlators and operator mixing in N=4 superYang-Mills theory}",
    eprint = "hep-th/0208178",
    archivePrefix = "arXiv",
    reportNumber = "AEI-2002-061",
    doi = "10.1016/S0550-3213(02)01025-8",
    journal = "Nucl. Phys. B",
    volume = "650",
    pages = "125--161",
    year = "2003"
}

@article{Gaiotto:2024,
    author = "Gaiotto, Davide and L{\'o}pez-Raven, Adri{\'a}n and Silverans, Hanne and Zeng, Keyou",
    title = "{Categorical {\textquoteright}t Hooft expansion and chiral algebras}",
    eprint = "2411.00760",
    archivePrefix = "arXiv",
    primaryClass = "hep-th",
    doi = "10.1007/JHEP12(2025)141",
    journal = "JHEP",
    volume = "12",
    pages = "141",
    year = "2025"
}

@article{KD,
    author = "Kontsevich, Maxim",
    title = "{Deformation quantization of Poisson manifolds. 1.}",
    eprint = "q-alg/9709040",
    archivePrefix = "arXiv",
    doi = "10.1023/B:MATH.0000027508.00421.bf",
    journal = "Lett. Math. Phys.",
    volume = "66",
    pages = "157--216",
    year = "2003"
}

@article{singleP,
    author = "Aprile, F. and Drummond, J. M. and Heslop, P. and Paul, H. and Sanfilippo, F. and Santagata, M. and Stewart, A.",
    title = "{Single particle operators and their correlators in free $ \mathcal{N} $ = 4 SYM}",
    eprint = "2007.09395",
    archivePrefix = "arXiv",
    primaryClass = "hep-th",
    doi = "10.1007/JHEP11(2020)072",
    journal = "JHEP",
    volume = "11",
    pages = "072",
    year = "2020"
}

@article{IKKT,
    author = "Ishibashi, N. and Kawai, H. and Kitazawa, Y. and Tsuchiya, A.",
    title = "{A Large N reduced model as superstring}",
    eprint = "hep-th/9612115",
    archivePrefix = "arXiv",
    reportNumber = "KEK-TH-503",
    doi = "10.1016/S0550-3213(97)00290-3",
    journal = "Nucl. Phys. B",
    volume = "498",
    pages = "467--491",
    year = "1997"
}

@article{Joao1,
    author = "Komatsu, Shota and Martina, Adrien and Penedones, Joao and Vuignier, Antoine and Zhao, Xiang",
    title = "{Einstein gravity from a matrix integral -- Part II}",
    eprint = "2411.18678",
    archivePrefix = "arXiv",
    primaryClass = "hep-th",
    month = "11",
    year = "2024"
}

@article{Joao2,
    author = "Komatsu, Shota and Martina, Adrien and Penedones, Jo{\~a}o and Vuignier, Antoine and Zhao, Xiang",
    title = "{Einstein gravity from a matrix integral -- Part I}",
    eprint = "2410.18173",
    archivePrefix = "arXiv",
    primaryClass = "hep-th",
    month = "10",
    year = "2024"
}

@article{Sean1,
    author = "Hartnoll, Sean A. and Liu, Jun",
    title = "{The polarised IKKT matrix model}",
    eprint = "2409.18706",
    archivePrefix = "arXiv",
    primaryClass = "hep-th",
    doi = "10.1007/JHEP03(2025)060",
    journal = "JHEP",
    volume = "03",
    pages = "060",
    year = "2025"
}

@article{Bonelli,
    author = "Bonelli, Giulio",
    title = "{Matrix strings in pp wave backgrounds from deformed superYang-Mills theory}",
    eprint = "hep-th/0205213",
    archivePrefix = "arXiv",
    reportNumber = "ULB-TH-02-15",
    doi = "10.1088/1126-6708/2002/08/022",
    journal = "JHEP",
    volume = "08",
    pages = "022",
    year = "2002"
}

@article{hoppe,
  title={QUANTUM THEORY OF A MASSLESS RELATIVISTIC SURFACE AND A TWO-DIMENSIONAL BOUND STATE PROBLEM},
  author={Jens Hoppe},
  journal={Soryushiron Kenkyu Electronics},
  volume={80},
  number={3},
  pages={145-202},
  year={1989},
  doi={10.24532/soken.80.3_145}
}

@article{KKN,
    author = "Kazakov, Vladimir A. and Kostov, Ivan K. and Nekrasov, Nikita A.",
    title = "{D particles, matrix integrals and KP hierarchy}",
    eprint = "hep-th/9810035",
    archivePrefix = "arXiv",
    reportNumber = "HUTP-98-A051, ITEP-TH-35-98, CERN-TH-98-302, SACLAY-SPH-T-98-102, LPTENS-98-40",
    doi = "10.1016/S0550-3213(99)00393-4",
    journal = "Nucl. Phys. B",
    volume = "557",
    pages = "413--442",
    year = "1999"
}

@article{Anderson:2016rcw,
    author  = "Anderson, Peter D. and Kruczenski, Martin",
    title   = "{Loop Equations and bootstrap methods in the lattice}",
    journal = "Nucl. Phys. B",
    volume  = "921",
    pages   = "702--726",
    year    = "2017",
    eprint  = "1612.08140",
    archivePrefix = "arXiv",
    primaryClass  = "hep-th"
}

@article{Lin:2020mme,
    author  = "Lin, Henry W.",
    title   = "{Bootstraps to strings: solving random matrix models with positivity}",
    journal = "JHEP",
    volume  = "06",
    pages   = "090",
    year    = "2020",
    eprint  = "2002.08387",
    archivePrefix = "arXiv",
    primaryClass  = "hep-th"
}

@article{Avan:1991kq,
  author  = "Avan, Jean and Jevicki, Antal",
  title   = "{Classical integrability and higher symmetries of collective string field theory}",
  journal = "Phys. Lett. B",
  volume  = "266",
  pages   = "35--41",
  year    = "1991"}

@article{Jha,
    author = "Jha, Raghav Govind",
    title = "{Introduction to Monte Carlo for matrix models}",
    eprint = "2111.02410",
    archivePrefix = "arXiv",
    primaryClass = "hep-th",
    doi = "10.21468/SciPostPhysLectNotes.46",
    journal = "SciPost Phys. Lect. Notes",
    volume = "46",
    pages = "1",
    year = "2022"
}

@article{Das:1991qb,
  author  = "Das, Sumit R. and Dhar, Avinash and Mandal, Gautam and Wadia, Spenta R.",
  title   = "{Gauge theory formulation of the c = 1 matrix model: Symmetries and discrete states}",
  journal = "Int. J. Mod. Phys. A",
  volume  = "7",
  pages   = "5165--5196",
  year    = "1992",
  eprint  = "hep-th/9110021",
  archivePrefix = "arXiv"}

@article{Dhar:1992hr,
  author  = "Dhar, Avinash and Mandal, Gautam and Wadia, Spenta R.",
  title   = "{Classical Fermi fluid and geometric action for c=1}",
  journal = "Int. J. Mod. Phys. A",
  volume  = "8",
  pages   = "325--350",
  year    = "1993",
  eprint  = "hep-th/9204028",
  archivePrefix = "arXiv"}

@article{Filev:2013pza,
    author  = "Filev, Veselin G. and O'Connor, Denjoe",
    title   = "{Multi-matrix models at general coupling}",
    journal = "J. Phys. A",
    volume  = "46",
    pages   = "475403",
    year    = "2013",
    eprint  = "1304.7723",
    archivePrefix = "arXiv",
    primaryClass  = "hep-th"
}

@article{Kazakov:2021lel,
    author  = "Kazakov, Vladimir and Zheng, Zechuan",
    title   = "{Analytic and numerical bootstrap for one-matrix model and ``unsolvable'' two-matrix model}",
    journal = "JHEP",
    volume  = "06",
    pages   = "030",
    year    = "2022",
    eprint  = "2108.04830",
    archivePrefix = "arXiv",
    primaryClass  = "hep-th"
}

@article{Berenstein:2008eg,
    author = "Berenstein, David E. and Hanada, Masanori and Hartnoll, Sean A.",
    title = "{Multi-matrix models and emergent geometry}",
    eprint = "0805.4658",
    archivePrefix = "arXiv",
    primaryClass = "hep-th",
    reportNumber = "NSF-KITP-08-68, WIS-10-08-MAY-DPP",
    doi = "10.1088/1126-6708/2009/02/010",
    journal = "JHEP",
    volume = "02",
    pages = "010",
    year = "2009"
}

@article{GMV,
    author = "Guerrieri, Andrea and Murali, Harish and Vieira, Pedro",
    title = "{Universality of heavy operators in matrix models}",
    eprint = "2507.21207",
    archivePrefix = "arXiv",
    primaryClass = "hep-th",
    doi = "10.1007/JHEP02(2026)074",
    journal = "JHEP",
    volume = "02",
    pages = "074",
    year = "2026"
}

@article{Vescovi:2024fwt,
    author = "Vescovi, Edoardo and Zarembo, Konstantin",
    title = "{Loop equations for generalised eigenvalue models}",
    eprint = "2402.13835",
    archivePrefix = "arXiv",
    primaryClass = "hep-th",
    doi = "10.21468/SciPostPhys.17.1.017",
    journal = "SciPost Phys.",
    volume = "17",
    number = "1",
    pages = "017",
    year = "2024"
}

@article{Budzik:2023xbr,
    author = "Budzik, Kasia and Gaiotto, Davide and Kulp, Justin and Williams, Brian R. and Wu, Jingxiang and Yu, Matthew",
    title = "{Semi-Chiral Operators in 4d ${N}=1$ Gauge Theories}",
    eprint = "2306.01039",
    archivePrefix = "arXiv",
    primaryClass = "hep-th",
    month = "6",
    year = "2023"
}
\appendix 
\setcounter{secnumdepth}{3}
\section{Scaling, Commutativity, and Axial Potentials} \label{appA}
\subsection*{Scaling and Universality}
In \cite{GMV} it was observed that the strong coupling limit $g\to \infty$ of the conventionally normalized deformed Hoppe model as $Z=\int dX dY \exp(-N S)$, with 
\beq
\!\!S=\tr\Big(\!-\tfrac{g^2}2 [X,Y]^2+\tfrac{1}{2}X^2+\tfrac{1}{2}Y^2+\sum_{n,m} t_{n,m}X^n Y^m\Big) , \label{A1}
\eeq
is universal in the sense that the deformation with $O(1)$ times $t_{n,m}$ with $n+m\geq3$ have no effect. Namely, the eigenvalue densities from which we can compute any disk correlators are simply given by the vacuum model with $t_{n,m}=0$ for which the eigenvalue densities are explicitly given by 
\beqa
&&\rho(x) = \frac{3(\ell^2 - x^2)}{4\ell^3} \, ,\qquad \ell = \left(\frac{3\pi}g\right)^{\frac13}\,.\label{rho1}
\eeqa
Furthermore, changing the quadratic term or introducing more general sources (like stacks of determinants, etc.) merely rescales this same density. Instead, our model (\ref{ZgenH}) amounts to scaling the times with appropriate powers of $g$ such that they matter and backreact on these densities. In an effective field theory analogy, we could think of (\ref{A1}) as an EFT with higher and higher terms being more and more suppressed at low energies while (\ref{ZgenH}) would correspond to a high energy situation where all EFT terms need to be taken into account and lead to genuinely new effective backgrounds. 
\\~\\
\subsection*{Commutativity} The generalized Hoppe model is commuting at strong coupling, as mentioned in the main text. To see why, note that we obtain the following loop equation by rescaling $X$ and $Y$ by some constant factor:
\beq
    2N=\left\langle\tr\left[\frac1g\left(X:\partial_xH:+Y:\partial_yH:\right)
 -4g\,[X,Y]^2\right]\right\rangle .
\label{eq:HoppeCommutatorDilation}\nn
\eeq
The moments of $X$ and $Y$ remain finite in the scaling used in
(\ref{ZgenH}), and hence
\beq
 \left\langle\tr [X,Y]^2\right\rangle
 =-\frac{N}{2g}+O(Ng^{-2})\,.
\label{eq:HoppeCommutatorNorm}
\eeq

Now, consider a swap of two neighboring $X$ and $Y$, which gives
\beq
 \tr(PXYQ-PYXQ)=\tr(QP\,[X,Y])\,.
\eeq
Cauchy--Schwarz, applied to both the matrix trace and the positive matrix integral, gives \cite{Martina:2026qxg}
\beq
 \left|\left\langle\tr(QP\,[X,Y])\right\rangle\right|^2
 \leq
 \left\langle\tr\!\left[QP(QP)^\dagger\right]\right\rangle
 \left\langle-\tr [X,Y]^2\right\rangle .
\label{eq:HoppeOrderingCS}
\eeq
The first factor has the usual $O(g^0)$ scaling because the matrices have $O(1)$ eigenvalue support. The second is $O(1/g)$ by \eqref{eq:HoppeCommutatorNorm}. Thus changing the ordering inside a trace costs at least a factor $g^{-1/2}$ relative to a generic trace. Since this argument holds in the presence of a general source, differentiating the one-point result with respect to that source gives the same conclusion for all connected correlators.
\\~\\
\subsection*{Axial Potentials} 
When $H(X,Y)=Y^2+V(X)$ the theory (\ref{ZgenH}) can be solved following almost verbatim the vacuum solution of the Hoppe model. We integrate out the $Y$'s which are Gaussian and find the saddle point equation for the $X$ eigenvalues \cite{Hoppe,GMV}
\beq
 \frac{V'(x)}g= 2\fint dx'\, \frac{1}{(x-x')(1+g^2(x-x')^2)} \,\rho(x') 
\eeq
which can be solved for any $g$ following \cite{Hoppe}. At strong coupling the right hand side becomes simply $
-\tfrac{2\pi}g \rho'(x)$ so that the density is given by 
\beq
\rho_V(x)=\frac1{2\pi}(\Lambda_V - V(x))
\eeq
where the constant $\Lambda_V$ is fixed by the normalization condition 
\beq
    \int_{V(x) < \Lambda_V} dx\,\rho_V(x) = 1\,.\nonumber
\eeq

The joint density for the axial potential can be found by computing systematically the $Y$ Wick contractions, as explained in \cite{GMV} and reviewed in Appendix \ref{Brute}, leading to 
\beq
\rho_V(x,y)= \frac1{\pi^2}\sqrt{\Lambda_V - V(x) - y^2} \label{rho2}
\eeq
which indeed coincides with our main result (\ref{rhoMain}) when $H(X,Y)=Y^2+V(X)$. For the particular case of the vacuum, i.e. $H(x,y) = x^2 + y^2$, the density is given by
\beq
    \rho(x,y) = \frac1{\pi^2}\sqrt{L^2-x^2-y^2}\,,\quad L\equiv \left(\frac{3\pi}2\right)^{\frac13}
\eeq

To derive the more general result (\ref{rhoMain}), however, we clearly need new tools since all manipulations above relied heavily on integrating out the $Y$'s which we can only do when they enter in a Gaussian way.

\section{Two Area-Preserving Symmetries} \label{ClarificationAppendic}
Recall that any matrix model has a $U(N)$ redundancy under rotating all
matrices at once. This redundancy is a gauge symmetry: it preserves all
eigenvalues and so on. 

At large $N$ it can be geometrized and
related to an area preserving diffeomorphism, as explained by Hoppe in
\cite{Hoppe}, where he was studying the quantization of two dimensional
membranes in four dimensional space-time. 
Our area preserving symmetry here is quite different. It acts on the two
matrices $X,Y$ themselves, it is a genuine statement about the matrix model
rather than a redundancy. Being a redundancy, the $U(N)$ symmetry gives nothing
new: the loop equation (\ref{zero}) generated by $\delta X_i=[\varepsilon,X_i]$
is trivially satisfied, since the change of the measure and the change of the
action cancel term by term. Ours instead implies (\ref{FSym}) and hence our main
result (\ref{rhoMain}). It also acts on the matrix eigenvalues in the simplest
way imaginable -- by moving them around the plane -- and it is present for this
precise two matrix model and at strong coupling only. 

In Hoppe's construction
the area preserving symmetry
corresponds to reparametrizations of a membrane worldvolume, while ours acts on
the target $(x,y)$ and relates different theories. In fancier terms:
thinking of the eigenvalues as defining a map $\sigma \mapsto
\big(x(\sigma),y(\sigma)\big)$ from a worldvolume into the $(x,y)$ plane,
Hoppe's diffeomorphisms act by \textit{pre}-composition on the domain, while
ours act by \textit{post}-composition on the target. 

A final point of clarification: Hoppe's membrane analysis led to a matrix quantum mechanics; the zero dimensional model studied here appears in his section C.III of the same work \cite{Hoppe} as a separate calculation, interesting in its own right, which gives no information about the spectrum of that quantum mechanics.

\section{Weyl Ordered Identities and Details} \label{WeylAp}
Given a function $h(x,y)$ given by a sum of monomials such as $h=x^2 y$ we can define a fully symmetric matrix made out of $X$ and $Y$ as in (\ref{hNO}) or, more generally
\beq
\!
\!:\!x^n y^m\!:\,=\!\frac{n!\,m!}{(n+m)!}\times
\left(\!
\begin{gathered}
    \texttt{\small sum over all distinct orderings}\\
    \texttt{\small of $n$ $X$'s and $m$ $Y$'s}
\end{gathered} \!
\right)\nn \!.
\eeq
Sometimes, an equivalent definition is more useful. We first define $\widehat h$ as the Fourier transform of the function $h$
\beq
h(x_1,x_2) = \int d^2k\, \widehat h(k) \,e^{i k\cdot x} \,.
\eeq
Then 
\beq
:h(x_1,x_2): =\int d^2k\, \widehat h(k) \,e^{i k\cdot X}
\eeq
This is sometimes also called the Weyl ordering prescription. The equivalence of the two definitions follows easily once we observe that $e^{i k\cdot X}$ automatically generate all possible symmetric polynomials so that 
\beq
:e^{i k\cdot x}:\,= e^{i k\cdot X}
\eeq
This Fourier representation using $e^{ik\cdot X}$ can be quite useful because we have good control over derivatives of exponentials through the so called Duhamel formula, 
\beq
\frac{d}{dt} e^{A(t)}= \int_{0}^1 ds \,e^{A(t)s} A'(t) e^{A(t)(1-s)} \,.
\eeq
For example, the action of the derivative on $V$ in (\ref{zero}) can be easily shown to be equal to 
\beq
\left(\frac{\partial}{\partial X_i}\right)^{\!a}_{\,\,b} \!\!\!(V_i)^b_{\,\,a} =\!i\int \!d^2 k\, k^i\widehat{v}_i(k) \int\limits_0^1 ds \tr(e^{is k\cdot X})\tr(e^{i(1-s) k\cdot X})\nn
\eeq
but this vanishes since the area preserving condition (\ref{div}) in Fourier space is nothing but $k^i\widehat{v}_i(k)=0$. 

Next we have the action of $\tr( V \partial/\partial X)$ on the commutator square in $S$ leading to 
\beq
\tr [V^i,X^j][X_i,X_j]\,. \label{intermediate}
\eeq
Using $V^i =\epsilon^{ij}:\partial_j \phi:$ (note that in two dimensions (\ref{div}) is equivalent to the statement that $v^i=\epsilon^{ij}\partial_j \phi$), we see that (\ref{intermediate}) is basically given by $\tr\Big( [:\partial_j \phi:,X^j] [X^1,X^2] \Big)$
which vanishes since 
\beq
[:\partial_j \phi:,X^j]=0 \label{id1}
\eeq
for any $\phi$. The proof of this simple identity is again straightforward if we use the Fourier representation. 

Finally, to study the action of the derivative on the potential term we can make use of the nice property
\begin{equation}
\frac{\partial}{\partial X_i}
\tr(: h :)= :\frac{\partial}{\partial x_i} h :\label{id2}
\end{equation}
which can again be established effortlessly using the Fourier representation. Using this we immediately get to the main result (\ref{product}) in the main text.

Let us conclude this appendix with a check of the two main identities (\ref{id1}), (\ref{id2}) which the reader already established in full generality using the Fourier representations. Let us consider some simple function $\phi=x^2 y$ so that (\ref{id1})
\begin{eqnarray*}
[:\partial_j \phi:,X^j] &=& [:2 x y:,X] +[:x^2:,Y] \nn \\
&=& [XY+YX,X]+[X^2,Y]=0 \,\,\,\checkmark \nn
\end{eqnarray*}
as it ought to. Note that the result is zero due to a nice cancellation between the two terms. Similarly, we can check (\ref{id2}) for $h=x^2 y$ say,
\begin{eqnarray*}
\frac{\partial}{\partial X} \tr:x^2y: &=& \frac{\partial}{\partial X} \tr (\tfrac{1}{3}X^2Y+\tfrac{1}{3}XYX+\tfrac{1}{3}YX^2) \nn \\
&=&XY+YX\nn\\
&=&:2xy:\nn\\
&=&:\frac{\partial}{\partial x} x^2 y: \,\,\, \checkmark \nn
\end{eqnarray*}

\section{Distributional Densities and Moments} \label{wmomentaAp}
The vertices $\omega_m$ used in the main text are derivatives of the density with respect to the normalization parameter $\Lambda$. The subtlety is that the support of the density also moves with $\Lambda$. It is therefore important to define these derivatives after doing the integral. For a fixed background $H_0$, let
\beq
 \rho_\Lambda(\vec x)=\frac1{\pi^2}
 \big(\Lambda-H_0(\vec x)\big)_+^{1/2}
\eeq
and let $\Lambda_0$ be fixed by $\int d^2x\,\rho_{\Lambda_0}=1$. We define $\omega_m$ by its integral against any smooth function $f$,
\beq
\!\! \int d^2x\,\omega_m(\vec x)f(\vec x)
 \!\equiv\!
 \left.
 \frac{\partial^{m-1}}{\partial\Lambda^{m-1}}
 \!\int d^2x\,\rho_\Lambda(\vec x)f(\vec x)
 \right|_{\Lambda=\Lambda_0}\!\!\! .
\label{eq:omegaDistributionDefinition}
\eeq
Here, the derivative is not allowed to pass through the integral without also keeping track of the moving boundary.

Let us now specialize to the vacuum, i.e. $H_0 = x^2 + y^2$ and $\Lambda_0 = L^2$. The lowest vertex $\omega_2^{(0)}$ is still an ordinary integrable function,
\beq
 \omega_2^{(0)}(\vec x)
 =\frac{\Theta\big(L^2-|\vec x|^2\big)}
 {2\pi^2\sqrt{L^2-|\vec x|^2}}.
\label{W2}
\eeq
From which we can compute the propagator $\kappa^{(0)}=\frac\pi L$. For $m\geq3$, the kernels are singular at the boundary and cannot be treated as ordinary functions. We can instead compute the moments of these vertices, which is all we need in practice. The RHS of \eqref{eq:omegaDistributionDefinition} before differentiating is given by
\beq
\begin{aligned}
 \int d^2x\,\rho^{(0)}_\Lambda(\vec x)|\vec x|^{2s}
 &=\frac{\Lambda^{s+3/2}}{2\sqrt\pi}
 \frac{\Gamma(s+1)}{\Gamma(s+5/2)}.
\end{aligned}
\label{rhoM}
\eeq
Taking $m-1$ derivatives with respect to $\Lambda$ and then setting $\Lambda=L^2$ gives
\beq
 \int d^2x\,\omega_m^{(0)}(\vec x)|\vec x|^{2s}
 =\frac{L^{5+2(s-m)}}{2\sqrt\pi}
 \frac{\Gamma(s+1)}{\Gamma(s-m+7/2)},
\eeq
which is \eqref{eq:momentsW}.

\section{The $XYZ$ model } \label{xyzAp}

In this appendix, we derive an exact finite coupling Ward identity using three-dimensional divergence-free flows \eqref{eq:TwoWeylWard}. This was independently discovered by Davide Gaiotto. We then show that at strong coupling, this implies a volume preserving diffeomorphism invariance 
which allows us again to compute infinitely many connected correlators.

\subsection*{The Model is Not Commutative}
The $XYZ$ model \eqref{ZgenH3} has a cubic interaction term and a Weyl ordered potential $H(X,Y,Z)$. The cubic term is imaginary for Hermitian matrices, so the three-matrix measure is not positive which has important implications. 

Indeed, let us begin with the axial potential
\beq
 H_V=V(X)+Y^2+Z^2\,.
\eeq
We can integrate out $Z$ by writing
\beq
 Z=W+g[X,Y]\,.
\eeq
The action then becomes
\beq
 \frac 1g\tr W^2-g\tr[X,Y]^2
 +\frac 1g\tr\big(V(X)+Y^2\big)\,.
\eeq
Thus we recover the Hoppe model with an axial potential after integrating
out $Z$. 

However, as alluded to in the main text, this three matrix model is not fully commuting. The Cauchy--Schwarz argument for strong coupling commutativity from Appendix \ref{appA} does not apply because the measure is not positive. 

Let's look at a simple example of non-commutativity. Consider the following one point function in the vacuum, i.e. $H_0 = x^2+y^2+z^2$
\beq
    \langle\tr XYZ\rangle = g \langle\tr XY[X,Y]\rangle = \frac g2 \langle\tr[X,Y]^2\rangle\nn
\eeq
On the other hand, we also have $\langle\tr XZY\rangle = -\frac g2\langle\tr [X,Y]^2\rangle\neq \langle\tr XYZ\rangle$. Clearly, the model is far from commutative.

However, we can still define a density for fully symmetric traces, which we turn to now.

\subsection*{Volume Preserving Symmetry}

It is useful to work in a Fourier basis and define
\beq
 G_V(\vec k)=\frac1N\left\langle
 \tr e^{i(k_xX+k_yY+k_zZ)}\right\rangle_V\,,
\label{xyzFourierBasis}
\eeq
which is a generating function for symmetrized traces. The XYZ model with an axial potential is invariant under rotations of $Y$ and $Z$, and hence
\beq
 G_V(k_x,k_y,k_z)
 =G_V\left(k_x,\sqrt{k_y^2+k_z^2},0\right)\,.
\eeq
With this, it is easy to see that the density for symmetrized traces in the axial model is precisely the rotationally invariant uplift of \eqref{rho2}
\beq
    \rho_V(x,y,z)=\frac1{2\pi^2} \Theta\big(\Lambda-V(x)-y^2-z^2\big)\,.
\label{xyzAxialDensity}
\eeq
In particular, $V(x)=x^2$ gives the uniform ball in 3D.

To find the density in the presence of arbitrary Weyl ordered potential $H(x,y,z)$, we follow very similar logic as in the Hoppe model -- we deform the variables as $\delta X_i = :v_i(X):$, where $v$ is again a divergence free vector field. This already ensures that the Jacobian variation vanishes. Now consider the variation of the cubic term. It is convenient to work in Fourier modes. Let $v_a = \hat v_a \exp(i \vec k.\vec x) $. Divergence-free condition tells us $\hat v(k) = \vec k \times \vec w$ for some $\vec w$. Therefore,
\beq
\begin{aligned}
    \delta \tr X_1[X_2,X_3] &= \frac12\epsilon_{abc}\tr:v_a:[X_b,X_c]\\
    &= \tr e^{i\vec k\cdot\vec X}
 [\vec k\cdot\vec X,\vec w\cdot\vec X]=0\,,
\end{aligned}
\eeq
so once again, the only surviving term is the variation of the potential, which gives the finite coupling Ward identity
\beq
    \langle\tr:v_i::\partial_iH:\rangle = 0\,.\label{eq:TwoWeylWard}
\eeq
Suppose for the moment that, inside the expectation value, we can combine
the two Weyl orderings at leading order. The Ward identity would then become
\beq
 \left\langle\tr:v_i\partial_iH:\right\rangle_H=0\,,
 \qquad \partial_iv_i=0\,.\label{eq:OneWeylWard}
\eeq
This is precisely the Ward identity for volume preserving diffeomorphisms.
Under the same one-droplet assumptions as in the two-matrix model, we can
therefore map the axial potential to a general potential. Since the axial
density is uniform, the density for symmetrized traces must be
\beq
 \rho(\vec x)=\frac1{2\pi^2}
 \Theta\big(\Lambda-H(\vec x)\big)\,,
\label{xyzGeneralDensity}
\eeq
where $\Lambda$ is fixed by $\int d^3x\,\rho(\vec x)=1$. This is the main result of this appendix. 

We can use it for the purpose of computing symmetrized trace observables, provided  (\ref{eq:TwoWeylWard}) can indeed be simplified to (\ref{eq:OneWeylWard}). Next we check that this is the case. 

\subsection*{(For some observables) the Model is  Commutative}
It remains to justify the ordering step. For two polynomials $f$ and $h$,
define
\beq
 D[f,h]=\tr\left(:f::h:-:fh:\right)\,.
\label{xyzOrderingDefect}
\eeq
We will now use another loop equation to show that this term is subleading. Once again, it is enough to show this in Fourier basis
\beq
 f=e^{i\vec k\cdot\vec x}\,,\qquad
 h=e^{i\vec q\cdot\vec x}\,,
\eeq
It is straightforward to show that
\beq
\begin{aligned}
 D_{\vec k,\vec q}
 &\equiv \tr\left(
 e^{i\vec k\cdot\vec X}e^{i\vec q\cdot\vec X}
 -e^{i(\vec k+\vec q)\cdot\vec X}
 \right)\\
 &=\tr F_{\vec k,\vec q}
 [i\vec k\cdot\vec X,i\vec q\cdot\vec X]\,.
\end{aligned}
\label{xyzOneCommutator}
\eeq
where
\beq
 F_{\vec k,\vec q}=\int_0^1ds\int_0^sdt\,
 e^{i(s-t)\vec q\cdot\vec X}
 e^{i(1-s)(\vec k+\vec q)\cdot\vec X}
 e^{is\vec k\cdot\vec X}e^{it\vec q\cdot\vec X}\,.\nn
\eeq
Now consider the variation
\beq
 \delta X_a=U_a\,,\qquad
 U_a=-(\vec k\times\vec q)_aF_{\vec k,\vec q}\,.
\eeq
The key observation is then that the variation of the cubic term under this is precisely
\beq
 \delta_U\tr X_1[X_2,X_3]=D_{\vec k,\vec q}\,,
\eeq
and therefore the full loop equation for $\delta X_i = U_i$ takes the form
\beq
    \left\langle D_{\vec k,\vec q}\right\rangle
 =\frac1{2g}\left\langle\tr U_a:\partial_aH:\right\rangle\,.
\label{xyzSecondWard}
\eeq
The right-hand side is explicitly free of commutators and is $O(g^{-1})$ in our normalization. Therefore, at leading order, \eqref{eq:TwoWeylWard} is the same as \eqref{eq:OneWeylWard} \footnote{Note also that this statement is not the same as saying the model is commuting and does not contradict the earlier finding $\langle\tr XYZ\rangle\neq \langle\tr XZY\rangle$.}.

\subsection*{Feynman Diagrams and Correlators}
Following the Hoppe analysis, we can readily write the free energy for the 3D model as well as 
\beq
\mathcal{F}=\underset{\Lambda}{\operatorname{ext}} \left\{\Lambda- \frac{1}{2\pi^2}\!\int d^3x \,\big(\Lambda-H(\vec{x})\big)\, \Theta\big( \Lambda-H(\vec{x})\big)
\right\}\,. \nn
\eeq
Equivalently, since the density is a step rather than a power,
\beq
\mathcal{F}[H]=\underset{\substack{0\le\rho\le 1/2\pi^2\\ \int d^3x\,\rho=1}}{\operatorname{min}}\ \int d^3x\, H(\vec{x})\,\rho(\vec{x})\,.
\eeq
From these we can derive general moments as in the text for the pure $XYZ$ model. As emphasized below (\ref{eq:HoppeGeneralNTreeFormula}) the structure of the main formula (\ref{eq:HoppeGeneralNTreeFormula}) holds here as well since as emphasized below that formula, this structure was a direct consequence of volume-preserving symmetry which is present here as well! The only thing that changes are the dynamics of the model which are encoded in vertices 
\beq
\begin{aligned}
\omega_m(\vec{x})&\equiv
\left.\frac{\partial^{m-1}}{\partial\Lambda^{m-1}}
\frac{1}{2\pi^2}\Theta\big(\Lambda-H_0(\vec{x})\big)
\right|_{\Lambda=\Lambda_0}\\
&=\frac{1}{2\pi^2}\delta^{(m-2)}
\big(\Lambda_0-H_0(\vec{x})\big)\,.
\end{aligned}
\label{eq:xyzContactKernels}
\eeq
For the vacuum, $H_0 = x^2+y^2+z^2$, the moments explicitly take the form
\beq
    \int d^3x\, \omega_m(\vec x) |\vec x|^{2s}= \frac{L^{2(s-m)+5}}{\pi}\frac{\Gamma\!\left(s+\tfrac32\right)}{\Gamma\!\left(s-m+\tfrac72\right)}\,,
\eeq
where again $L=\left(\frac{3\pi}2\right)^{\frac13}$.

A particularly interesting set of fully symmetrized correlators are the chiral primary correlators (the name is purely by analogy with $\mathcal{N}=4$ SYM).  We take the polarizations to be null, $\vec n_i^{\,2}=0$, and define
\beq
G_n \equiv \left\<\prod_{i=1}^n \tr(n_i\cdot X)^{J_i}\right\>_c \,.
\eeq
For $n=2$ and $n=3$ only contact diagrams exist -- see figure \ref{fd_fig}. Let
\beq
\begin{aligned}
    G_2 &=\left(\delta_{J_1,J_2}\,d_{12}^{J_1}\right)\,\times\mathcal N_1\,,\\
    G_3&=\left(\frac{\sqrt{\mathcal N_1\mathcal N_2\mathcal N_3}}{N}\,d_{12}^{\ell_{12}}d_{13}^{\ell_{13}}d_{23}^{\ell_{23}}\right)\,\times c_{123}\,.
\end{aligned}
\eeq
where the \textit{propagators} $d_{ij}=\vec n_i\cdot\vec n_j$ and the \textit{bridge lengths}
\beq
    \ell_{ij}=\frac{J_i+J_j-J_k}{2}\,.
\eeq
At weak and strong coupling, the 2pt normalization is
\beq
\mathcal N_i=
\begin{cases}
\displaystyle J_i\left(\frac g2\right)^{J_i}\,,&g\ll1\,,\\[6pt]
\displaystyle g\,\frac{2^{J_i}(J_i!)^2}{(2J_i+1)!}
\frac{L^{2J_i+1}}{\pi}\,,&g\gg1\,,
\end{cases}
\eeq
while the normalized three point function is 
\beq
\begin{aligned}
\left.c_{123}\right|_{g\ll1}
&=\sqrt{J_1J_2J_3}\,,\\
\left.c_{123}\right|_{g\gg1}
&=\frac{\sqrt g}{\sqrt6\,L}\,
\binom{S/2}{\ell_{12},\ell_{13},\ell_{23}}\,
\frac{\sqrt{\prod_{i=1}^3(2J_i+1)!}}{S!}\,.
\end{aligned}  \nn
\eeq
where in the second line $S=J_1+J_2+J_3$. 
The leading order weak coupling result is the same as the corresponding correlators of CPOs in $\mathcal N=4$ SYM. Understanding the interpolation between the weak and strong coupling limit would be very instructive.

\section{Brute Force} \label{Brute}
In this appendix, we compute the two and three point functions of the Hoppe model in a much less elegant, but straightforward manner. Historically, this is how we first computed these correlators and only later found the powerful area preserving diffeomorphism symmetry. This appendix therefore serves as an independent check of the Witten-like diagrams of figure \ref{fd_fig}. 

The way we will proceed is to diagonalize $X$ and integrate out the $Y$ matrices by Wick contractions. It is useful to first put back the factors which were suppressed around \eqref{higherPt}. We define the leading planar correlators by
\beq
 \left\langle\prod_{i=1}^p\tr:f_i(X,Y):\right\rangle_c
 =\left(\frac g N\right)^{p-2}\,\mathcal C_p[f_1,\ldots,f_p]+\ldots \,.\nn
\eeq
such that for $H_t=x^2 + y^2 -\sum_i t_i f_i$, we have
\beq
\!\! \mathcal C_p[f_1,\ldots,f_p]
 =\!g\!\left.\frac{\partial}{\partial t_2}\cdots
 \frac{\partial}{\partial t_p}
 \int d^2x\,\rho_t(\vec{x})f_1(\vec{x})\right|_{t=0} .\la{higherPtApp}
\eeq

\subsection*{$X$-only correlators}
Let  us first compute the two and three point correlators that involve only the $X$ matrix. Recall that with an axial potential $V_t(x) = x^2 - \sum t_n x^n$, the eigenvalue density of $X$ is given by
\beq
    \rho(x;t) = \frac1{2\pi} (\Lambda(t) - V_t(x)) \Theta(\Lambda(t) - V_t(x))
\eeq
where we explicitly added the step function to emphasize that the derivatives in \eqref{higherPtApp} can act on it.  Let the support of the density be given by $x\in (a(t), b(t))$. The parameters $\Lambda(t), a(t)$ and $b(t)$ are then fixed by the following conditions
\begin{equation}
\begin{aligned}
    1&=\frac{1}{2\pi}\int_{a(t)}^{b(t)}dx\,
    \left(\Lambda(t)-x^2+\sum_{n\geq0}t_nx^n\right),
    \\
    \Lambda(t)&=V_t(a(t)) = V_t(b(t))\,.
\end{aligned}
\label{eq:bruteAxialNormalizationEndpoints}
\end{equation}
Without sources, the density is given by the parabola
\beq
    \rho(x;t=0) = \frac{3(L^2-x^2)}{4L^3}\,\Theta(L-|x|)\,, \label{eq:vacuumDensityBrute}
\eeq
where $L=\left(\frac{3\pi}{2}\right)^{1/3}$. We can solve for $\partial_{t_1}\ldots \partial_{t_p}\rho(x;t)$ by perturbatively expanding \eqref{eq:bruteAxialNormalizationEndpoints}. With a single derivative, the result is
\begin{equation}
   \left.\partial_{t_m}\rho(x; t)\right|_{t=0}
    \!=\!\frac{\Theta(L-|x|)}{2\pi}\left[
    x^m-\tfrac{L^m}{m+1}\tfrac{1+(-1)^m}{2}
    \right]
\label{eq:bruteMarginalFirstVariation}
\end{equation}
It turns out to be more convenient to define a density $\rho_X^{(p)}$ for connected higher point correlators
\beq
    \mathcal C_p[x^{n_1}, \ldots, x^{n_p}] = \int \prod_i (dx_i\, x_i^{n_i})\ \rho_X^{(p)}(x_1,\ldots x_p)\,.
\eeq
The first two axial connected densities are then given by

\begin{align}
    \rho_X^{(2)}(x_1,x_2) &= \frac g{2\pi}\Theta(L-|x_1|)\label{eq:twoPtXOnly}\\
    &\quad\times\left(\delta(x_1-x_2)-\frac1{2L}\Theta(L-|x_2|)\right)\nn\\
    \rho_X^{(3)}(x_1,x_2,x_3)&=\frac{3g}{8L}\sum_{\sigma=\pm1}\prod_{i=1}^3\left[\frac{\delta(x_i-\sigma L)}{L}-\frac{\Theta(L-|x_i|)}{2L^2}\right].\nn
\end{align}

\subsection*{$Y$ Wick contractions}
Now, let us look at correlators involving both $X$'s and $Y$'s.  We begin
with the one point function, which was computed in
\cite{GMV,Berenstein:2008eg}.  In an $X$ eigenvalue background, the $Y$
propagator in the normalization of \eqref{ZgenH} is
\begin{align}
 \left\langle Y_{ij}Y_{kl}\right\rangle_Y
 &=\frac{g}{2N}\frac{\delta_{il}\delta_{jk}}
 {1+g^2(x_i-x_j)^2}\,,\\
 &\underset{g\to\infty}{=}\frac\pi{2N}\delta_{il}\delta_{jk}\delta(x_i-x_j)\,,
\label{eq:bruteYPropagatorStrong}
\end{align}
where in the second line we used that the limit of the Lorentzian is a delta function. Consequently, all the index sums collapse into a single integral and the location of the $X$'s in the sea of $Y$'s is irrelevant (note that the $X$'s are diagonal and do not contribute any extra index sums; they simply live on the index lines, or faces, of the $Y$'s). Every planar contraction of $2s$ $Y$'s on a disk gives the same factor $(\pi/2)^s$.  The number of such planar contractions is counted by the Catalan numbers $C_s$. We have,
\begin{align}
 \frac1N\left\langle\tr X^nY^{2s}\right\rangle
 &=\frac1g\mathcal C_1[x^ny^{2s}]\nn\\
 &=C_s\left(\frac\pi2\right)^s
 \int_{-L}^{L}dx\,x^n\rho_X(x)^{s+1}.
\label{eq:bruteYOnePointMoments}
\end{align}
The odd moments vanish.  Here $\rho_X(x)=\rho(x;t=0)$~\eqref{eq:vacuumDensityBrute}. From this, the two dimensional density for one point functions immediately follows
\beq
    \rho^{(1)}(x,y) = \frac1{\pi^2}\sqrt{L^2-x^2-y^2}\ \Theta(L^2-x^2-y^2)\,,\nn
\eeq
in agreement with \eqref{rhoMain}.

\begin{figure}
    \centering
    \includegraphics[width=\linewidth, trim={12cm 15cm 6cm 0},clip]{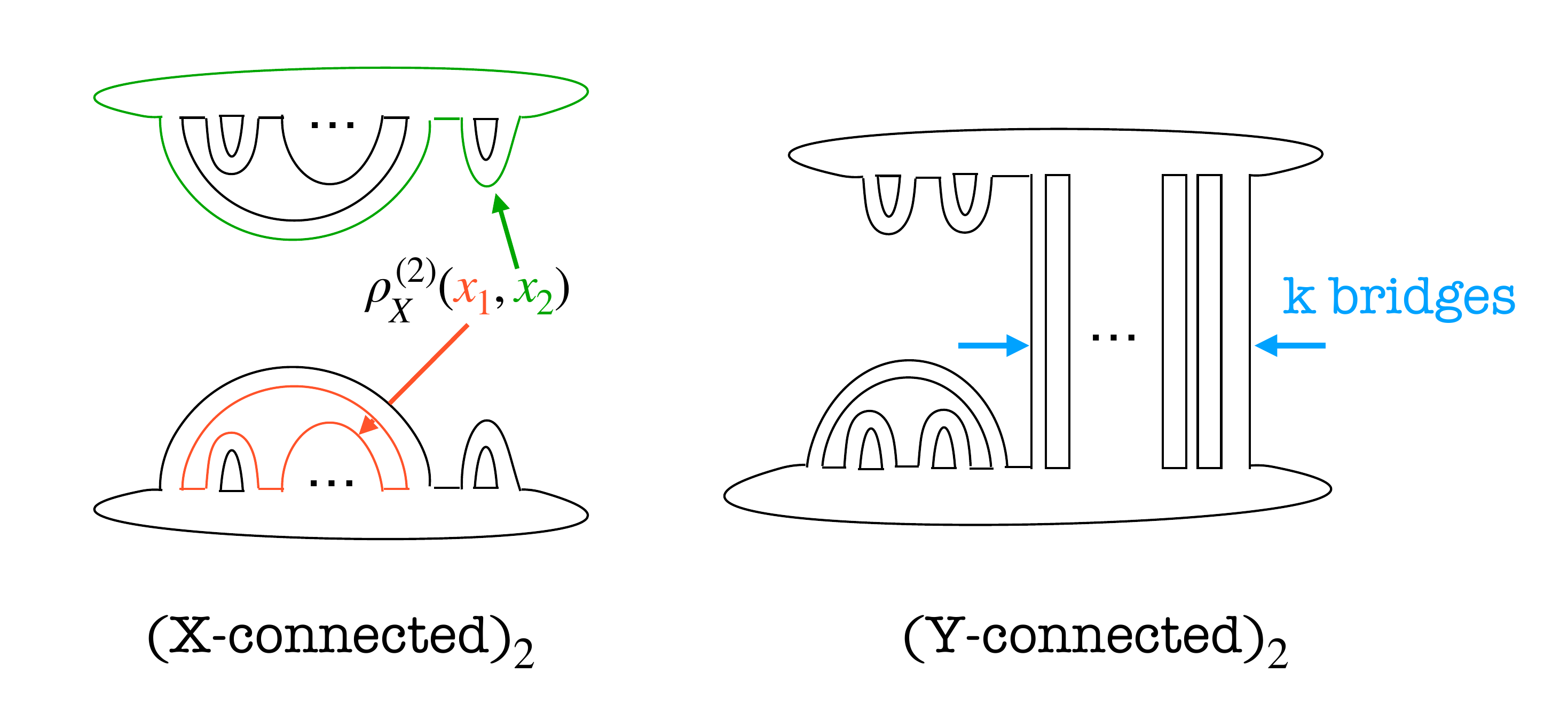}
    \caption{The two classes of planar $Y$ Wick contractions contributing to the connected two point function. On the left, all the $Y$'s are contracted within their own trace and the two disks are connected by $\rho_X^{(2)}$. On the right, the two traces are joined directly by $k$ $Y$ propagators, which produce the closed chain of Lorentzians in \eqref{eq:bruteLorentzianCycle}. The remaining $Y$'s are contracted within each trace. We separated the rainbows and the bridges for clarity, but in general they can interleave.}
    \label{fig:Brute2pt}
\end{figure}

Let us now move on to the two point function.  There are two classes of planar $Y$ contractions, see figure \ref{fig:Brute2pt}.  In the first, all the $Y$'s are contracted within their own trace.  The two traces must then be connected by an $X$ density fluctuation $\rho_X^{(2)}$ \eqref{eq:twoPtXOnly}.  For the pair of monomials $x^{n_i}y^{m_i}$, each disk has $\frac{m_i}2+1$ eigenvalue loops to which this fluctuation can attach.  This gives
\begin{align}
 \texttt{(X-connected)}_2
 ={}&\prod_{i=1}^2\left(\frac{m_i}2+1\right)\left(\frac\pi2\right)^{\frac{m_i}2}C_{\frac{m_i}2} \times \nn\\
 &\!\!\!\!\!\!\! \!\!\!\!\!\!\! \!\!\!\!\!\!\! \!\!\!\!\!\!\! \!\!\!\!\!\!\! \!\!\!\!\!\!\! \times\int dx_1dx_2\,\rho_X^{(2)}(x_1,x_2)
 \prod_{i=1}^2x_i^{n_i}\rho_X(x_i)^{\frac{m_i}2}.
\label{eq:bruteXTwoPointZeroBridge}
\end{align}
This term of course only exists when the $m_i$'s are even. In the second class of $Y$ contractions, the two traces are joined by $k\geq1$ $Y$ propagators. Here one cannot replace each Lorentzian separately by a delta function, since these terms contain a closed chain of Lorentzians, see figure \ref{fig:Brute2pt}. A chain of $k$ such Lorentzians obeys, with $x_{k+1}=x_1$,
\beq
 \prod_{r=1}^{k}\frac1{1+g^2(x_r-x_{r+1})^2}
 \to \frac1k\prod_{r=1}^{k-1} \frac\pi g\delta(x_r-x_{r+1}) \,.
\label{eq:bruteLorentzianCycle}
\eeq
Note that we have an extra $\frac1k$ compared to the naive rule of replacing each Lorentzian by a delta function (this would also have given a divergent $\delta(0)$). Now, we need to count the number of planar contractions on a cylinder with $k$ propagators going between the two traces with $m_1$ and $m_2$ $Y$'s respectively. This is given by
\beq
 A_{k;m_1,m_2}
 =k\binom{m_1}{(m_1-k)/2}\binom{m_2}{(m_2-k)/2},
\label{eq:bruteAnnularFixedBridges}
\eeq
where $1\leq k\leq\min(m_1,m_2)$ and $k$ has the same parity as both $m_1$ and $m_2$. Now since we have at least one $Y$ contraction, we do not need to use $\rho_X^{(2)}$. Once again, the delta functions collapse all index sums and we obtain
\begin{align}
 \texttt{(Y-connected)}_2
 ={}&
 \frac g\pi\left(\frac\pi2\right)^{\frac{m_1+m_2}2}
\sum_{k=1}^{\min(m_1,m_2)}\frac{A_{k;m_1,m_2}}k\times \nn\\[-1mm]
 &\!\!\!\!\!\!\times\int_{-L}^{L}dx\,
 x^{n_1+n_2}\rho_X(x)^{\frac{m_1+m_2}2}.
\label{eq:bruteYTwoPointAnnularMoments}
\end{align}
Finally, combining \eqref{eq:bruteXTwoPointZeroBridge} and \eqref{eq:bruteYTwoPointAnnularMoments} into a single two point density, we obtain
\begin{align}
\!\!\!\! \rho^{(2)}(\vec x_1,\vec x_2) = g\,\omega_2(\vec x_1)
\!\left[
\delta^{(2)}(\vec x_1-\vec x_2)-\tfrac{2L^2}3\omega_2(\vec x_2)
\right]\!
\label{eq:bruteYTwoPointDensity}
\end{align}
where we used the distributional density $\omega_2$ defined in~(\ref{eq:HoppeGeneralNLambdaRecursion}), (\ref{eq:momentsW}) or (\ref{W2}). Integrating against two wave-packets~$f_{i=1,2}(x,y)$, this precisely reproduces the result in the first line of figure \ref{fd_fig}. Note that the second term in the square brackets in (\ref{eq:bruteYTwoPointDensity}) translates nicely into the subtractions leading to the~$f\to h$ wave packet re-centering described in section \ref{secApplication}.

In a similar fashion, we can also compute the connected three point functions. In principle we would need to repeat the counting exercise keeping track of how many bridges we have between the three traces. However, there is a simpler way using rotational invariance: we will first compute $\langle\tr X^{n_1}Y^{m_1}\tr X^{n_2}Y^{m_2}\tr X^{n_3}\rangle_c$, and then uplift this to a rotationally invariant result. This way, we only need the cylinder combinatorics \eqref{eq:bruteAnnularFixedBridges}.

There are again two classes of $Y$ contractions. First consider contractions which are disconnected in $Y$, so that $m_1$ and $m_2$ must both be even. After contracting away the $Y$'s, the three traces can be connected directly by $\rho_X^{(3)}$. This gives
\begin{align}
 \texttt{(X-connected)}_3
 ={}&\prod_{i=1}^2C_{\frac{m_i}{2}} \left(\frac{m_i}{2}+1\right)\left(\frac\pi2\right)^{\frac{m_i}{2}}\label{eq:bruteYThreePointXBlock}\\
 &\times\int dx_1dx_2dx_3\,
 \rho_X^{(3)}(x_1,x_2,x_3)\nn\\
 &\times\rho_X(x_1)^{\frac{m_1}{2}}
 \rho_X(x_2)^{\frac{m_2}{2}}
 \prod_{i=1}^3x_i^{n_i}.\nn
\end{align}
Instead of using a $\rho_X^{(3)}$ to connect the traces, we can also use two copies of $\rho_X^{(2)}$ to connect them in a tree. We use one of the traces as a base and connect one eigenvalue-loop each with the other two traces. We obtain
\begin{align}
 \texttt{(X-tree)}_3
 ={}&\frac1g\prod_{i=1}^2C_{\frac{m_i}{2}} \left(\frac{m_i}{2}+1\right)\left(\frac\pi2\right)^{\frac{m_i}{2}}\\
 & \times\int \prod_{i=1}^3 \left(dx_i\, x_i^{n_i}\rho_X(x_i)^{\frac{m_i}2}\right)\,\rho_X^{(2)}(x_1,x_2)\nn\\
 &\hspace{0.5cm}\times
 \left[
 \frac{m_1\rho_X^{(2)}(x_1,x_3)}{2\rho_X(x_1)}
 +\frac{m_2\rho_X^{(2)}(x_2,x_3)}{2\rho_X(x_2)}
 \right]\,,\nn
\label{eq:bruteYThreePointXTree}
\end{align}
where $m_3$ is understood to be zero in the second line.

Finally, the first two traces can be connected by $Y$ propagators. If $k$ propagators run between them, the same annular counting as in the two point function applies. The resulting cylinder has $(m_1+m_2)/2$ eigenvalue loops, and one of them must connect to the third trace through $\rho_X^{(2)}$. Thus
\begin{align}
 \texttt{(Y-connected)}_3
 ={}&\frac{s}{\pi}
 \left(\frac\pi2\right)^{s} \left(\sum_{k=1}^{\min(m_1,m_2)}
 \frac{A_{k;m_1,m_2}}k\right)\nn\\
 &\!\!\!\!\!\!\!\!\!\!\!\!\!\!\!\!\!\!\times\int dx\,dx_3\,
 \rho_X(x)^{s-1}
 \rho_X^{(2)}(x,x_3)\nn x^{n_1+n_2}x_3^{n_3}.\nn
\end{align}
Here $m_1$ and $m_2$ can be either both even or both odd and we defined $s=\frac{m_1+m_2}2$. The three point function with the third trace containing only $X$'s is the sum of the two \texttt{connected} terms and the \texttt{tree} term.

Although the third trace was treated differently in this calculation, the answer has a rotationally invariant uplift. In practice, to find this uplift we first computed $\langle\tr Z^{n_1}\bar Z^{m_1}\tr Z^{n_2}\bar Z^{m_2} \tr X^{n_3}\rangle_c$ where $Z=X+i Y$. The $U(1)$ rotational symmetry allows us to replace the last trace by $\tr (Z\bar Z)^{\frac{n_3-J}2}Z^J$ for an appropriate $J$. We skip these details here for brevity. The resulting three point density $\rho^{(3)}(\vec{x}_1,\vec{x}_2,\vec{x}_3)$ is given by
\beq
    g\!\int \!d^2u\,\omega_3(\vec{u})\prod_{i=1}^3\left[
 \delta^{(2)}(\vec{x}_i-\vec{u})-\tfrac{2L^2}3\omega_2(\vec{x}_i)
 \right]\,,\nn
\eeq
which is strikingly simpler than any intermediate result! 
This again matches \eqref{eq:HoppeGeneralNTreeFormula} -- or second term in figure \ref{fd_fig} -- after integrating against test functions. (Again, note the nice emergence of the $f\to h$ re-centering of the wave packets by the $\omega_2$ terms in this final expression.)

With some more blood, sweat, and tears, we can  compute the four point density in the same way. Of course, the final result agrees with the much simpler derivation in the main text. 
\end{document}